%% file: main.tex
\documentclass[pdflatex,sn-mathphys-num,iicol]{sn-jnl}

\usepackage{booktabs}
\usepackage{multirow}
\usepackage{mathtools}
\usepackage{xcolor}
\usepackage{gensymb}
\usepackage{subfig}
\usepackage{siunitx}
\usepackage{chemformula}
\usepackage{amssymb}
\usepackage{lineno}

\input{mydef}

\begin{document}

\title{Ultra-trace analysis of U and Th in organic liquid scintillators with high sensitivity}

\author*[a]{A.~Barresi}\email{andrea.barresi@unimib.it}
\author[a]{D.~Chiesa}
\author[b]{D.~Merli}
\author*[a]{M.~Nastasi}\email{massimiliano.nastasi@unimib.it}
\author[c]{S.~Nisi}
\author[a,c]{E.~Previtali}
\author[a]{M.~Sisti}
\author[a]{M.~Borghesi}
\author[d,m]{A.~Cammi}
\author[a]{C.~Coletta}
\author[a]{G.~Ferrante}
\author[d]{L.~Loi}
\author[e]{G.~Andronico}
\author[f]{V.~Antonelli}
\author[f]{D.~Basilico}
\author[f]{M.~Beretta}
\author[g]{A.~Bergnoli}
\author[f]{A.~Brigatti}
\author[g]{R.~Brugnera}
\author[e]{R.~Bruno}
\author[h]{A.~Budano}
\author[f]{B.~Caccianiga}
\author[g]{V.~Cerrone}
\author[e]{R.~Caruso}
\author[i]{C.~Clementi}
\author[g]{L.V.~D'Auria}
\author[g]{S.~Dusini}
\author[h]{A.~Fabbri}
\author[j]{G.~Felici}
\author[g]{A.~Garfagnini}
\author[f]{M.G.~Giammarchi}
\author[e]{N.~Giudice}
\author[g]{A.~Gavrikov}
\author[g]{M.~Grassi}
\author[e]{N.~Guardone}
\author[f]{F.~Houria}
\author[f]{C.~Landini}
\author[g]{L.~Lastrucci}
\author[g]{I.~Lippi}
\author[f]{P.~Lombardi}
\author[k,l]{F.~Mantovani}
\author[h]{S.M.~Mari}
\author[j]{A.~Martini}
\author[f]{L.~Miramonti}
\author[k,l]{M.~Montuschi}
\author[h]{D.~Orestano}
\author[i]{F.~Ortica}
\author[j]{A.~Paoloni}
\author[f]{L.~Pelicci}
\author[f]{E.~Percalli}
\author[h]{F.~Petrucci}
\author[f]{G.~Ranucci}
\author[f]{A.C.~Re}
\author[k,l]{B.~Ricci}
\author[i]{A.~Romani}
\author[f]{P.~Saggese}
\author[g]{A.~Serafini}
\author[g]{C.~Sirignano}
\author[g]{L.~Stanco}
\author[h]{E.~Stanescu Farilla}
\author[k,l]{V.~Strati}
\author[f]{M.D.C~Torri}
\author[e]{C.~Tuve'}
\author[h]{C.~Venettacci}
\author[e]{G.~Verde}
\author[j]{L.~Votano}

\affil[a]{\orgname{INFN Sezione di Milano Bicocca e Dipartimento di Fisica, Università di Milano Bicocca}, \orgaddress{\city{Milano}, \country{Italy}}}
\affil[b]{\orgname{Department of Chemistry, University of Pavia}, \orgaddress{\city{Pavia},  \country{Italy}}}
\affil[c]{\orgname{INFN, Laboratori Nazionali del Gran Sasso}, \orgaddress{\city{Assergi}, \country{Italy}}}
\affil[d]{\orgname{INFN Sezione di Milano Bicocca e Dipartimento di Energetica, Politecnico di Milano}, \orgaddress{\city{Milano}, \country{Italy}}}
\affil[e]{\orgname{INFN Sezione di Catania e Università di Catania, Dipartimento di Fisica e Astronomia}, \orgaddress{\city{Catania}, \country{Italy}}}
\affil[f]{\orgname{INFN Sezione di Milano e Università degli Studi di Milano, Dipartimento di Fisica}, \orgaddress{\city{Milano}, \country{Italy}}}
\affil[g]{\orgname{INFN Sezione di Padova e Università di Padova, Dipartimento di Fisica e Astronomia}, \orgaddress{\city{Padova}, \country{Italy}}}
\affil[h]{\orgname{INFN Sezione di Roma Tre e Università degli Studi Roma Tre, Dipartimento di Matematica e Fisica}, \orgaddress{\city{Roma}, \country{Italy}}}
\affil[i]{\orgname{INFN Sezione di Perugia e Università degli Studi di Perugia, Dipartimento di Chimica, Biologia e Biotecnologie}, \orgaddress{\city{Perugia}, \country{Italy}}}
\affil[j]{\orgname{INFN, Laboratori Nazionali di Frascati}, \orgaddress{\city{Frascati}, \country{Italy}}}
\affil[k]{\orgname{INFN, Sezione di Ferrara}, \orgaddress{\city{Ferrara}, \country{Italy}}}
\affil[l]{\orgname{Università degli Studi di Ferrara, Dipartimento di Fisica e Scienze della Terra}, \orgaddress{\city{Ferrara}, \country{Italy}}}
\affil*[m]{\orgname{Emirates Nuclear Technology Center (ENTC), Khalifa University} \orgaddress{\city{Abu Dhabi}, \country{United Arab Emirates}}}

\abstract{
Rare event searches demand extremely low background levels, necessitating ever-advancing screening techniques to enhance sensitivity. Liquid scintillators are highly attractive as detector media due to their inherent radiopurity and scalability in mass. In this work, we present a screening procedure to measure ultra-trace concentrations of natural contaminants -- \u\ and \th\ -- with sensitivities at the \qty{E-15}{g/g} level. Our method combines neutron activation analysis with radiochemical techniques, followed by \bg\ coincidence spectroscopy to minimize interference backgrounds. This approach achieves sensitivities of \qty{0.65E-15}{g/g} for \u\ and \qty{1.9E-15}{g/g} for \th, among the best reported worldwide. Potential pathways for further sensitivity improvements are outlined in the conclusions.
}


\maketitle

\section{Introduction}
\label{S:Introduction}
\input{1.Introduction}

\section{Neutron activation analysis sensitivity}
\label{S:NAA}
\input{2.NAA}

\section{Measurement Strategy}
\label{S:MeasStrategy}
\input{3.MeasurementStrategy}
       
    \subsection{Labware cleaning protocol}
    \label{S:Labware}
    \input{3.1.LabwareCleaning}

    \subsection{Radiochemical procedures}
    \label{S:RadioProcedure}
    \input{3.2.RadioChemProcedure}
    
    \subsection{Neutron irradiation facility}
    \label{S:Triga}
    \input{3.3.NeutronIrradiation}

    \subsection{The \bg\ system GeSparK}
    \label{S:GeSpark}
    \input{3.4.GeSpark}

\section{Recovery efficiency determination}
\label{S:RecoveryEff}
\input{4.RecoveryEfficiency}

\section{Measurement of blank samples}
\label{S:Blanks}
\input{5.BlankSamples}

\section{Conclusions}
\label{S:concl}
\input{Conclusions}

\begin{appendices}
\input{appendix_LOD}
\input{appendix_Np-239}
\end{appendices}

\bibliography{Bibliography}

\end{document}

%% file: mydef.tex
\def\u{$^{238}$U}
\def\th{$^{232}$Th}
\def\np{$^{239}$Np}
\def\pa{$^{233}$Pa}

\def\utevaR{UTEVA\textsuperscript{\textregistered}}
\def\uteva{UTEVA}
\def\tevaR{TEVA\textsuperscript{\textregistered}}
\def\teva{TEVA}
\def\truR{TRU\textsuperscript{\textregistered}}
\def\tru{TRU}
\def\bg{$\beta$/$\gamma$}
\def\g{$\gamma$}
\def\b-{$\beta^-$}
\def\ppq{\qty{E-15}{g/g}}
\def\be{\begin{equation}}
\def\ee{\end{equation}}

\def\ohm{$\Omega$}

%% file: 1.Introduction.tex
Rare event searches, particularly in the field of neutrino physics, require massive detectors with exceptionally low background event rates (see, e.g.,\cite{juno-bkg,borexino-nature,sno+,gerda,cupid-mo,cuore-bkg,legend-1000} for a non-exhaustive list of references). Background control has become the central challenge for experiments in this domain, making the careful selection of construction materials a key strategy in minimizing spurious events. This, in turn, places stringent demands on laboratory screening techniques, which must operate at the forefront of technological capabilities to validate materials with the highest sensitivity.

This paper presents a methodology for screening liquid scintillator matrices with sensitivities reaching the \ppq\ concentration level for natural radioactive contaminants such as \u\ and \th. The approach combines neutron activation analysis with radiochemical processing and utilizes a custom \bg\ coincidence detector \cite{GS:2021} to analyze the irradiated samples in a low-background configuration, maximizing detection sensitivity after neutron activation.
This method is specifically applied to a liquid alkyl benzene (LAB)-based scintillator, the primary solvent selected as the detector medium for the Jiangmen Underground Neutrino Observatory -- JUNO experiment \cite{juno-ppnp}, which requires extremely demanding radiopurity level \cite{juno-bkg} and optical properties \cite{refractiveIndex-juno,fluorescence-juno}, but it could also be adapted for use with other liquid scintillators in future experiments.

The paper is structured as follows: Section~\ref{S:NAA} provides an overview of neutron activation analysis and explores methods to enhance its sensitivity. Section~\ref{S:MeasStrategy} details our measurement strategy, outlining each step of the procedure. In Section~\ref{S:RecoveryEff}, we present our findings on the recovery efficiency of the entire process, while Section~\ref{S:Blanks} focuses on blank sample measurements and the achieved sensitivities for detecting trace concentrations of \u\ and \th. Finally, the Conclusions discuss potential paths for further improvements.

%% file: 2.NAA.tex
Neutron Activation Analysis (NAA) is a highly effective technique for achieving remarkable sensitivity in detecting stable or long-lived nuclides, such as $^{238}\textrm{U}$ and $^{232}\textrm{Th}$ \cite{NAA-2024}.
This method leverages the conversion of the target nuclides into short-lived isotopes with high specific activities, which can be readily measured using \g\ spectroscopy.

The activation reactions are radiative $(n,\gamma)$ captures on \u\ and \th,  generating radionuclides according to the following sequences:
\begin{align*}
    & ^{238}\text{U} \xrightarrow{(n,\gamma)} \, ^{239}\text{U} \xrightarrow[23.45\text{ m}]{\beta^-} \, ^{239}\text{Np} \xrightarrow[2.356\text{ d}]{\beta^-} \, ^{239}\text{Pu}\\
    & ^{232}\text{Th} \xrightarrow{(n,\gamma)} \, ^{233}\text{Th} \xrightarrow[22.3\text{ m}]{\beta^-} \, ^{233}\text{Pa} \xrightarrow[26.97 \text{ d}]{\beta^-} \, ^{233}\text{U}\\
\end{align*}    
The activities of \np\ and \pa,  quantified through \g\ spectroscopy, provide the basis for determining the concentrations of the corresponding target elements.
For a sample $S$ with mass $m_S$ (e.g., the liquid scintillator mass), the concentration of the target nuclide $i$ (e.g., \u\ or \th) can be calculated as:
\begin{equation}
\label{Eq:concentrazione}
    C_i = \dfrac{1}{m_S} \, \dfrac{M_i \, \mathcal{N}_i}{N_\textrm{Av}} = \dfrac{M_i }{m_S\, N_\textrm{Av}}\dfrac{R_i}{\sigma_\text{eff} \, \Phi}
\end{equation}
Here $\mathcal{N}_i$ is the number of target nuclides of species $i$, $M_i$ is its isotopic mass, and $N_\textrm{Av}$ is Avogadro's number.
The parameters $\sigma_\text{eff}$ and $\Phi$ denote the effective activation cross section and the total neutron flux intensity, respectively, with their product typically determined using activation standards irradiated alongside the sample $S$.  
The activation rate $R_i$ is directly proportional to the number of decays from the activated nuclide recorded in the \g\ spectrum. This proportionality takes into account the detection efficiency, the duration of the irradiation and \g\ measurement, and the time elapsed between the end of irradiation and the spectroscopy measurement. For high-sensitivity measurements, achieving a high and uniform neutron flux is essential. For this reason, nuclear reactors are commonly employed as neutron sources.
More details about neutron activation analysis for radiopurity screening are discussed in \cite{NAA-2024}.

By taking advantage of low-background High Purity Germanium (HPGe) detectors for measuring the activated sample,  sensitivities of the order of \qty{E-12}{g/g} are achievable \cite{NAA-2024}.
Enhancing sensitivity requires increasing the sample mass and minimizing residual background. As shown in Eq.~\ref{Eq:concentrazione}, in the absence of background, the measured concentration $C_i$ -- and consequently the NAA sensitivity -- scales linearly with the number of target nuclides in the sample $S$. However, the presence of background can interfere with the evaluation of the number of events at the photopeak of the activated nuclide. In such cases, only an upper limit on $C_i$ can be determined: 
\begin{equation}
    C_i^{\text{max}} \propto \sqrt{B \, \Delta E \, T}
\end{equation}
where $\Delta E$ is the energy range in which the peak is expected, $B$ is the background counting rate per unit energy, and $T$ is the measurement time. 
Improvements in sensitivity are then constrained, scaling only as the square root of the background reduction factor, as dictated by the Poissonian statistical fluctuations in the number of background events within the region of interest. A more detailed discussion on the evaluation of detection limits is provided in \ref{S:lod}.

Our strategy to increase the effective sample mass involves the use of radiochemical treatments. The actual sample mass is constrained by the maximum volume that can be accommodated in the reactor's irradiation channel. By employing a procedure to concentrate the target nuclides, we effectively increase the useful mass without exceeding the volume limit for neutron irradiation.
In addition, appropriate radiochemical treatments are specifically designed to remove interfering isotopes -- nuclides within the sample matrix that are activated by neutrons, producing nuclear decays that generate correlated background in \g\ spectroscopy measurements.
Finally, the measurement sensitivity can be further enhanced by employing a custom \bg\ coincidence detector, like the GeSpark system~\cite{GS:2021,GSMuons}. This detector leverages the coincidence between \b-\ particles detected by a liquid scintillation detector and \g\ rays detected by an HPGe detector, effectively reducing correlated and uncorrelated background by selecting \bg\ events corresponding to the beta decays of the activated nuclides of interest.

Our final procedure, designed to achieve sensitivity at the \qty{E-15}{g/g} level for \u\ and \th, combines NAA, radiochemical techniques, and \bg\ coincidence measurements. It is applicable to any sample composed of an apolar organic liquid matrix, which is a common solvent in the production of liquid scintillators (LS). The details are provided in the following section.

%% file: 3.MeasurementStrategy.tex
As outlined in the previous section, enhancing the sensitivity of NAA involves measures such as increasing the effective mass of the irradiated sample and eliminating interfering isotopes. In the case of LS, these two objectives can be achieved through chemical and radiochemical treatments that allow the sample to be concentrated while at the same time separating the isotopes of interest from interfering ones. 
Our finalized procedure follows the sequence outlined below, as illustrated in the block diagram of Figure \ref{fig:RadiochemicalPrinciple}:
\begin{figure*}
    \begin{center}
        \includegraphics[width=0.9\textwidth]{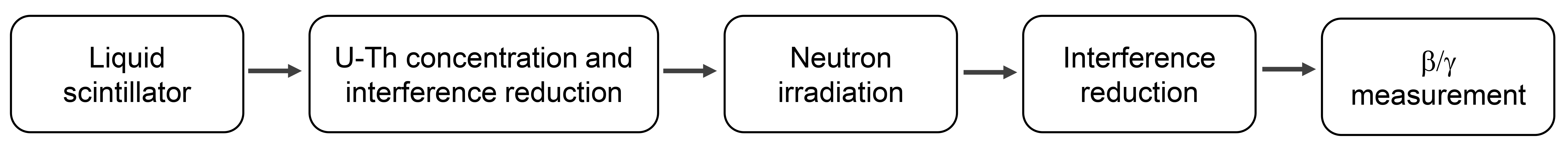}
        \caption{Block diagram of the measurement procedure.}
        \label{fig:RadiochemicalPrinciple}
    \end{center}
\end{figure*}
\begin{enumerate}
    \item Liquid-liquid extraction, which allows the transfer of the elements of interest (U and Th) from the organic matrix to an acidic aqueous solution;
    \item Pre-irradiation extraction chromatography with \utevaR resins, which enables the separation of different chemical species based on their chemical affinity and the acid concentration of the filtered solution;    
    \item Neutron irradiation at a nuclear reactor;
    \item Post-irradiation extraction chromatography with \tevaR resins, which are better-suited for separating neutron-activated nuclides such as \np\ and \pa;
    \item \bg\ spectroscopy, which exploits the coincident emission of beta and gamma radiations during the \b-\ decays of \np\ and \pa, to reduce the correlated and uncorrelated background.
\end{enumerate}
All processes, except for neutron irradiation, are conducted at the laboratories of the University of Milano-Bicocca.

A critical aspect of any analytical procedure for detecting trace concentrations of natural radioisotopes in samples is the implementation of stringent cleaning protocols to prevent the unintentional introduction of these contaminants prior to neutron irradiation. Additionally, the radiopurity of the reagents used in pre-irradiation treatments plays a pivotal role. Section~\ref{S:Labware} will provide a detailed discussion of these challenges. Section~\ref{S:RadioProcedure} will describe the radiochemical steps in our procedure. 
Section~\ref{S:Triga} will provide an overview of the irradiation facility, while Section~\ref{S:GeSpark} will briefly introduce the \bg\ spectroscopy system.

%% file: 3.1.LabwareCleaning.tex
The maximum allowable concentration of $^{238}\textrm{U}$ and $^{232}\textrm{Th}$ in LS, as required by the most recent rare event experiments (\qty{E-15}{g/g}), is several orders of magnitude lower than their typical concentration in the environment.
To achieve such ultra-low sensitivities, the cleanliness of the working environment, materials, consumables, and reagents is critical.

All treatments performed before sample irradiation must take place in a rigorously controlled, clean environment, with a very low concentration of airborne particulate. Specifically, these processes are conducted in a cleanroom classified as Class $10000$ (ISO7), meaning no more than \qty{10000}{particles/ft^3} greater than \qty{0.5}{\micro\m} are present. Radiochemical treatments are carried out under a laminar-flow hood, which improves cleanliness to Class 1000 (ISO 6) within the immediate working area.

One of the most critical sources of contamination is the reagents used during the radiochemical processes in the pre-irradiation phases. In particular, two reagents are of primary concern: water and nitric acid. Given the goal of achieving a measured sensitivity of \qty{E-15}{\g/\g}, the concentration of uranium and thorium in these reagents must be kept well below this threshold.

Water is purified using the Milli-Q or Milli-Q Element ultra-pure water systems from Merck Millipore, ensuring a conductivity of \qty{18.2}{\mega\ohm\cm} and no particulates greater than \qty{0.22}{\micro\m}. The Milli-Q Element system is specifically designed for trace-element analysis. Ultra-pure nitric acid from Carlo Erba Reagenti, with less than \qty{E-14}{g/g} uranium and thorium, is used and stored in PFA (Perfluoroalkoxy) bottles.

To verify the radiopurity of the reagents, samples of Milli-Q water, Milli-Q Element water, and ultra-pure nitric acid were sent to the Gran Sasso National Laboratory (LNGS) for ICP-MS analysis. For this purpose, PFA containers -- cleaned for several weeks with ultra-pure nitric acid -- were shipped from LNGS to Milano-Bicocca Laboratory, filled with the samples, and returned for screening.
ICP-MS measurements were performed using a sector field (SF)-ICP-MS (Finnigan Element 2, Thermo Scientific). To achieve the best sensitivity, the mass spectrometer was equipped with a ``Jet interface pump option'', X-cone type, and an Apex Q high-sensitivity introduction system (Element Scientific Inc.) which operates the vaporization chamber at the temperature of \qty{100}{\celsius} and the Peltier cell condenser at the temperature of \qty{2}{\celsius}, which are the optimal conditions for aqueous samples. In this setup, signal intensities above 10000 and 15000 cps (counts per second) were obtained for pg/g of \th\ and \u\, respectively.

The ICP-MS measurement results, shown in Table \ref{tab:ReagentValidation}, confirm the quality of the Milli-Q water, with uranium and thorium concentrations below \qty{E-15}{g/g}. No differences were observed between the Milli-Q and Milli-Q Element waters in terms of uranium and thorium content. The purity of the nitric acid was also confirmed, though the measured limit is slightly higher. This is due to the ICP-MS measurement limitations for nitric acid solutions with concentrations below 10\%, necessitating dilution with water, which reduced the sensitivity.
\begin{table}
    \centering
    \begin{tabular}{ccc}
        \toprule
        & $^{238}\textrm{U} (g/g)$ & $^{232}\textrm{Th} (g/g)$ \\
        \midrule
        \ch{H2O} Milli-Q & \num{<0.7E-15} & \num{<0.8E-15} \\
        \ch{H2O} Milli-Q Element & \num{<0.7E-15} & \num{<0.8E-15} \\
        Ultra-pure \ch{NHO3} & \num{<3E-14} & \num{<3E-14} \\
        \bottomrule
    \end{tabular}
    \caption{Results of the measurement performed at LNGS for the validation of the purity of the reagents used in the pre-irradiation treatments.}
    \label{tab:ReagentValidation}
\end{table}

\begin{figure}
   \centering
    {\includegraphics[width=.30\textwidth]{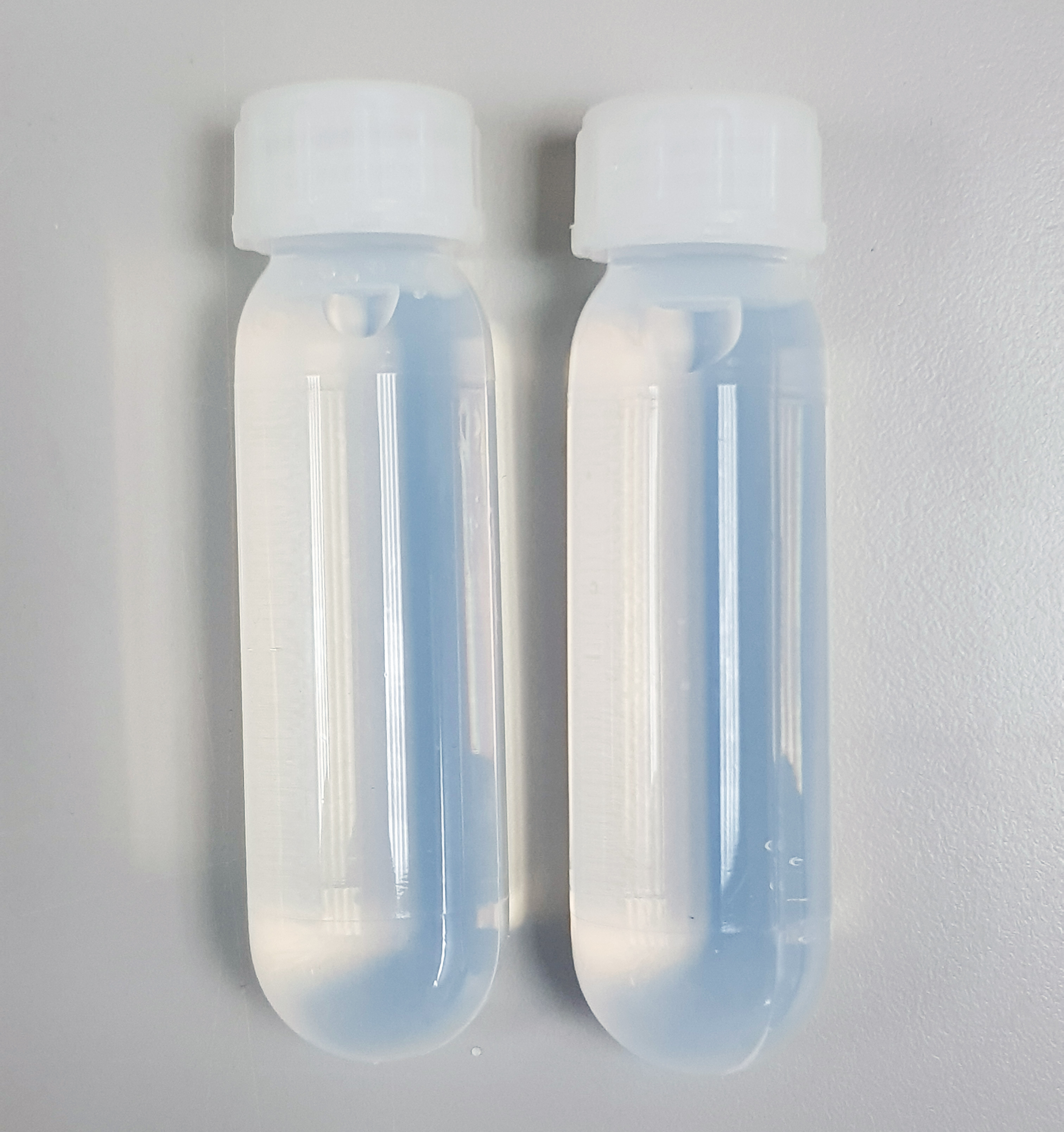}}
    \caption{PFA vials with screw caps.}
\label{fig:VialsPFA}
\end{figure}

Another critical issue concerns containers. Reagents used in the radiochemical treatments, as well as sample solutions, must be prepared in contamination-free containers, preventing exposure to external contaminants. Throughout our tests, we identified the best-performing containers for this purpose. Initially, we used 50 mL polypropylene centrifuge tubes (produced by Falcon or Corning) with screw caps, which were suitable for sealing long-term acid solutions. However, these tubes did not fit in the irradiation containers from the nuclear reactor because of their size.
We then tested polyethylene (PE) vials with pressure caps but encountered issues: handling the pressure caps introduced an additional risk of contaminating the liquid sample, and ICP-MS measurements at LNGS confirmed that the intrinsic contamination levels in these vials were higher than in other containers. ICP-MS analyses were carried out to quantify \u\ and \th\ concentrations in the cleaning solutions stored for at least one week in six containers of each type. The results, shown in Table \ref{tab:VialValidation}, demonstrated significant contamination and high variability in the cleaning solutions of PE vials, making them unsuitable for sample preparation. In contrast, we found that Falcon centrifuge tubes exhibited much lower and more reproducible contamination levels, making them suitable for temporary storage of radiochemical solutions.
For the final sample irradiation, we opted for PFA centrifuge vials with screw cap from Nalgene (see Figure \ref{fig:VialsPFA}), which fit the irradiation containers perfectly. PFA offers chemical inertness and resistance to strong acids and solvents and can be easily cleaned, making it ideal for ultra-trace element analysis. Table \ref{tab:VialValidation} presents the very good results we obtained for these containers using ICP-MS.

\begin{table}
    \centering
    \begin{tabular}{ccc}
        \toprule
        & $^{238}\textrm{U}$ & $^{232}\textrm{Th}$ \\
        & (\qty{E-15}{g/g}) & (\qty{E-15}{g/g}) \\
        \midrule
        PE vials        & \num{100(100)}$^{(\dagger)}$    & \num{180(70)}$^{(\dagger)}$  \\
        Falcon tubes    & \num{2.0(0.9)}                & \num{7.9(1.9)} \\
        PFA vials       & \num{1.7(0.8)}                & \num{2.6(1.4)} \\
        ICP-MS blank    & \num{0.6(0.4)}                & \num{0.6(0.3)} \\
        \bottomrule
    \end{tabular}
    \caption{Results of the measurements conducted at LNGS for the validation of the sample containers. The table presents the concentrations of \u\ and \th\ in the cleaning water after a two-week cleaning period. The errors are given as the standard deviation of the measurement results. $(\dagger)$ A high standard deviation indicates significant variability in the observed contamination, due to the high contamination risk associated with these vials.}
    \label{tab:VialValidation}
\end{table}

Finally, for liquid extraction and chromatography, we use ultra-pure nitric acid stored in \qty{500}{mL} PFA bottles. These bottles, which contain highly concentrated nitric acid, are particularly suitable for long-term storage and large-volume radiochemical processing, ensuring the highest level of cleanliness.

In our procedures, all items are thoroughly cleaned to reduce uranium and thorium release to minimize contamination from instruments and consumables. The cleaning protocol, conducted inside the clean room, uses a solution of ultra-pure water (Milli-Q) and 2\% ultra-pure nitric acid, which dissolves nuclides from material surfaces. 
Consumables like vials, bottles, and centrifuge tubes are immersed in this solution for several days, then filled with fresh acid and left for at least one week before use. If not used immediately, the solution is replaced every two weeks. 
Other items, such as micropipette tips, are cleaned for a week and then dried and stored under vacuum.

%% file: 3.2.RadioChemProcedure.tex
\subsubsection{Liquid-liquid extraction}
The goal of liquid-liquid extraction is to transfer a solute from one solvent to another. This is achieved by exploiting the different affinities of the solute for the two solvents, which must necessarily be immiscible with each other. In our case, the solvents are an apolar organic phase and an extracting aqueous polar phase consisting of a nitric acid solution, which extracts the metals to form water-soluble salts.
This procedure also allows for a pre-concentration of the sample, as the volume of the extracting solution is smaller than that of the sample. Additionally, it allows the sample to be treated with subsequent extraction chromatography steps, which cannot be performed with an organic matrix.
    
When the two immiscible solvents are in contact, solutes migrate from the organic phase to the aqueous phase until equilibrium is reached. This equilibrium is described by the distribution coefficient K, which is the ratio of the solute concentration in the organic phase to that in the aqueous phase.
Since K is constant for a specific combination of analytes, solvents, and experimental conditions -- such as temperature and pH -- and is independent of the solvent volumes, higher extraction efficiency can be achieved by performing the extraction process multiple times with small volumes of fresh extracting solution rather than using a single large volume.
Another important aspect is that the volume of the extracting solution cannot be too small to guarantee good contact between the two phases.
Typically, a volume ratio of total aqueous to organic solvents of 1:3 is used to concentrate the sample, and the extraction process is repeated three times for optimal efficiency.
This means that for \qty{100}{\mL} of organic solvent, we use a total extracting volume of \qty{30}{\mL} divided into three extractions with \qty{10}{\mL} for each one. Each extraction step lasts \qty{20}{minutes}, during which the mixture is vigorously stirred using a magnetic stirrer to ensure thorough contact between the two solvents.
The complete liquid-liquid extraction procedure is the following:
    \begin{enumerate}
        \item The extracting solution is prepared by diluting the nitric acid with water up to a concentration of \qty{1}{M}.
        \item The organic solvent and $1/3$ of the extracting solution are mixed and vigorously stirred by a magnetic stirrer for \qty{20}{minutes}.
        \item The mixture is transferred into a separatory PTFE (polytetrafluoroethylene) funnel, where the organic and the water phases are separated thanks to their immiscibility and different densities.
        \item The mixing and separation processes are repeated two additional times.
    \end{enumerate}
Starting with \qty{1}{L} of LS, the liquid-liquid extraction process results in approximately \qtyrange[range-units=single,range-phrase=-]{200}{300}{mL} of sample solution.

\subsubsection{Extraction chromatography}

\begin{figure}
        \begin{center}
            \includegraphics[width=0.3\textwidth]{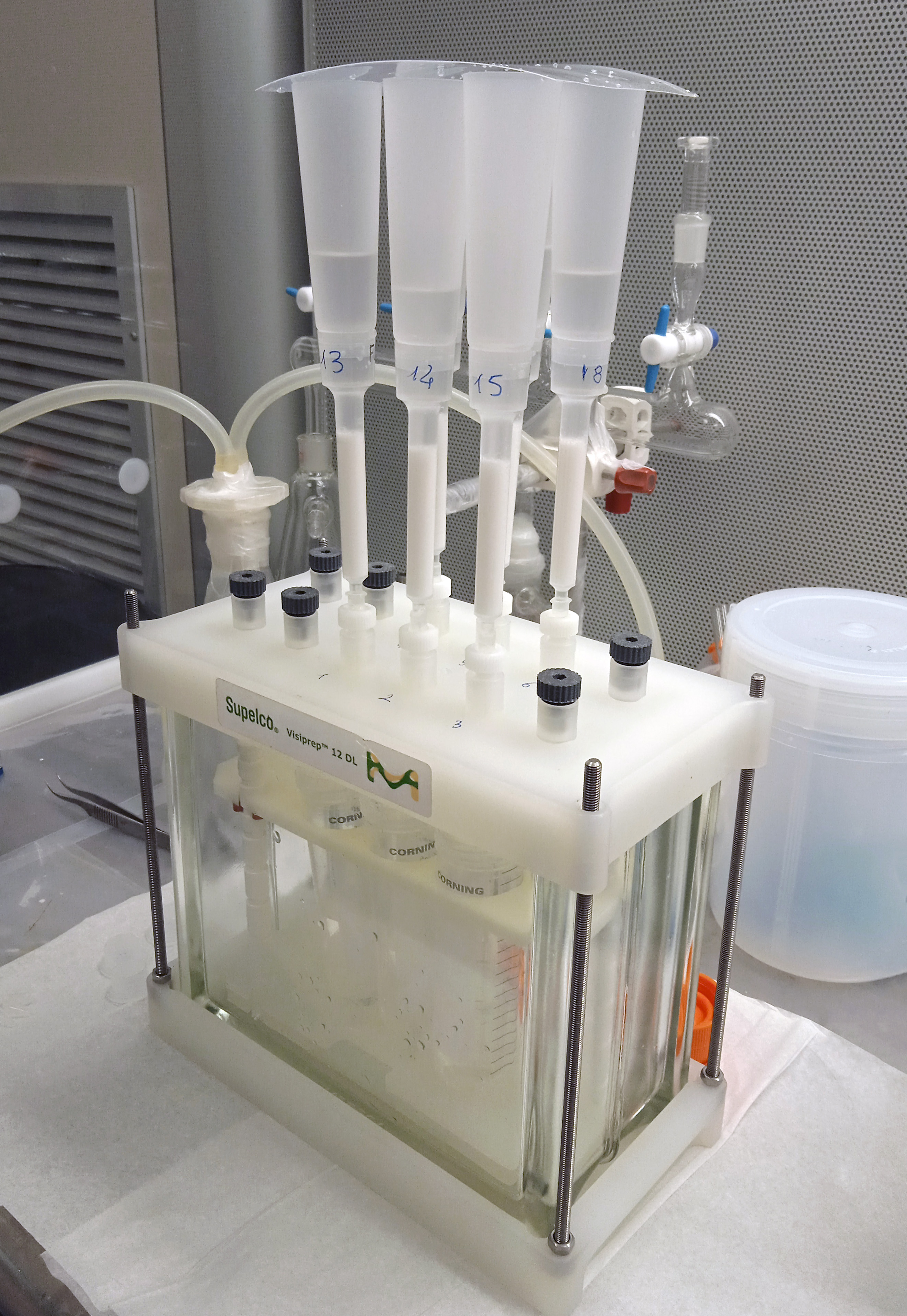}
            \caption{Vacuum system used for the extraction chromatography process. This system is necessary to guarantee a precise and stable flow rate of the solutions through the columns.}
            \label{fig:VacuumSystem}
        \end{center}
    \end{figure}

Extraction chromatography (EC) using resins is a separation technique in which the extractant is immobilized on an inert support through adsorption. The aqueous sample containing the analytes of interest flows through this resin. The extractant is highly selective for specific ions, and its affinity can be adjusted by modifying the pH or the composition of the solution being processed.
The resin is commercially available in pre-packed \qty{2}{\mL} columns equipped with a small funnel at the top to facilitate solution loading. A vacuum system is employed to ensure the proper flow rate of solutions through the column, as EC columns require a flow rate in the range of \qtyrange{0.6}{0.8}{\mL/\min}. Figure \ref{fig:VacuumSystem} provides an image of the setup.    
EC with column resins typically involves the following steps.
The first step is \textbf{preliminary washing}, in which the resin in the column is thoroughly washed with a dilute acid solution to remove the preservative liquid and any potential traces of target elements that may be present in the resin. This step is particularly important when analyzing very low concentrations, as even trace contaminants can affect the results, as discussed in Section~\ref{S:Blanks}. The dilute acid solution helps release any ions bound to the resin. 
Next, the resin is \textbf{conditioned} by flushing it with a solution having the same acid concentration as the sample. This prepares the resin to retain the target nuclides dissolved in the sample during the extraction process.
Then, the resin is \textbf{charged}, meaning that the sample is passed through the resin. During this process, the nuclides of interest are retained by the resin's stationary phase, while other components remain in the mobile phase, which is referred to as the ``leachate'' when it exits the column.
After that, there may be an additional \textbf{washing} phase, where the resin is rinsed again with an acid solution of the same concentration as the sample. This step removes any residual sample solution remaining in the column without extracting the nuclides of interest from the stationary phase.
Finally, the \textbf{elution} solution is passed through the column to release the nuclides of interest from the resin's stationary phase. The resulting solution is referred to as the ``eluate''.

Our measurement strategy involves two EC steps: one prior to neutron irradiation and one following it. The pre-irradiation step serves a dual purpose: reducing the sample mass and eliminating potential interfering nuclides that could be activated during the neutron irradiation, thereby minimizing correlated background. The post-irradiation step focuses on removing activated interfering nuclides to minimize the correlated background for the subsequent \g\ spectroscopy.

\paragraph{Pre-irradiation EC}
The \utevaR\ resin (Uranium and TEtraValents Actinides) is widely used for the separation of U and tetravalent actinides such as Th, Np, and Pu. We selected this resin for pre-irradiation treatment due to its exceptional efficiency in extracting $^{238}\textrm{U}$ and $^{232}\textrm{Th}$, the nuclides of interest in this stage. The protocol for \uteva\ treatment was developed based on manufacturer guidelines and literature references. This protocol was optimized to maximize recovery efficiency and applied to the sample following liquid-liquid extraction, prior to irradiation. Our final \uteva\ procedure is the following:
    \begin{enumerate}
        \item \textbf{Washing}: to ensure thorough cleaning of the resin, two washing steps are performed, each using \qty{50}{\mL} of \qty{0.02}{M} ultra-pure \ch{HNO3}, with the steps separated by the charging of \qty{10}{mL} of \qty{5}{M} ultra-pure \ch{HNO3} (conditioning).
        \item \textbf{Conditioning}: it is performed using \qty{15}{mL} of \qty{5}{M} ultrapure \ch{HNO3}.  It ensures that the resin is properly conditioned to the correct pH for optimal retention of the $^{238}\textrm{U}$ and $^{232}\textrm{Th}$ nuclides from the sample.
        \item \textbf{Charging}: the sample is passed through the resin with a \qty{5}{M} concentration  of \ch{HNO3}.
        \item \textbf{Elution}: it is performed using \qty{15}{mL} of \qty{0.02}{M} ultra-pure \ch{HNO3} to guarantee the complete release of the analytes of interest.
    \end{enumerate}
In this case, the additional washing step is omitted, as it is not critical prior to irradiation. Furthermore, it could increase the risk of analyte loss and sample contamination.
At the end of the pre-irradiation EC, the initial volume of \qtyrange[range-units=single,range-phrase=-]{200}{300}{mL} is further reduced to approximately \qty{15}{mL} of a mildly acidic solution, ready for neutron irradiation.

\paragraph{Post-irradiation EC}
The \tevaR\ resin (TEtraValents Actinides) is primarily utilized for the separation of tetravalent actinides. We chose this resin for post-irradiation treatments because, in this case, the goal is to extract  $^{239}\textrm{Np}$ and $^{233}\textrm{Pa}$, both of which are tetravalent actinides. \teva\ resin offers higher retention factors for these nuclides compared to \uteva, making it more suitable for this purpose.
Also for the \teva\ resin, the optimal protocol was developed based on manufacturer guidelines and literature references and optimized to maximize the recovery efficiency. This protocol was applied to the sample after the neutron irradiation.
Since $^{239}\textrm{Np}$ and $^{233}\textrm{Pa}$ can exist in multiple stable oxidation states, and the \teva\ resin selectively works with tetravalent elements (Np(IV) and Pa(IV)), it is crucial to ensure that the elements are in the correct oxidation state. This is achieved by adding a set of reagents to the sample to adjust the redox potential of the solution, thus ensuring the oxidation state remains IV. After neutron irradiation, the radiopurity of the reagents is no longer a concern, as the target elements have been converted into artificial nuclides. For the same reason, the preliminary washing step is omitted, since $^{239}\textrm{Np}$ and $^{233}\textrm{Pa}$ cannot be present in the resin. Our final \teva\ procedure is the following:
    \begin{enumerate}
        \item \textbf{Correct oxidation state}: after adjusting the concentration of \ch{HNO3} to \qty{8}{M}, the following reagents are added to the sample solution:
        \begin{itemize}
            \item \qty{25}{\micro\L} sulfamic acid (\ch{H2NSO3H}) \qty{1.5}{M};
            \item \qty{2}{\micro\L} \ch{Fe(NO3)3} \qty{5}{\mg/\mL} in \ch{HNO3} \qty{1}{M};
            \item \qty{75}{\micro\L} ascorbic acid (\ch{C6H8O6}) \qty{1}{M}.
        \end{itemize}
        These quantities are calculated for \qty{1}{mL} sample at the correct nitric acid concentration. The  volumes of reagents must be adjusted based on the sample volume at the required nitric acid concentration. Allow the sample to rest for at least \qty{5}{minutes} before charging it into the column. 
        \item \textbf{Conditioning}: it is performed using \qty{15}{mL} of \ch{HNO3} \qty{8}{M}. This step ensures that the resin has the correct pH for retaining the $^{239}\textrm{Np}$ and $^{233}\textrm{Pa}$ during extraction from the sample.
        \item \textbf{Charging}: while passing through the resin, the sample must have a concentration of \qty{8}{M} \ch{HNO3} and the correct redox potential.
        \item \textbf{Washing}: this step is performed with \qty{15}{\mL} of \ch{HNO3} \qty{8}{M} to remove residual traces of interfering nuclides.
        \item \textbf{Elution}: it is performed with \qty{15}{mL} of a solution containing: 
        \begin{itemize}
            \item \ch{HNO3} \qty{2}{M};
            \item Ammonium oxalate (\ch{(NH4)2C2O4}) \qty{0.5}{M};
            \item Hydroxylamine hydrochloride (\ch{NH2OH*HCl}) \qty{0.01}{M}.
        \end{itemize}
    \end{enumerate}	
After the post-irradiation EC, the final product is a mildly acidic solution with a volume of less than \qty{15}{mL}, ready to be placed in the GeSpark detector for \g\ spectroscopy.

%% file: 3.3.NeutronIrradiation.tex
For neutron irradiation, we utilize the TRIGA Mark II nuclear research reactor at the University of Pavia, located at about \SI{40}{km} from the University of Milano-Bicocca. This pool-type reactor, cooled and partially moderated by light water, operates at a maximum nominal power of \qty{250}{kW}. Its cylindrical core (approximately \qty{46}{cm} in diameter and \qty{36}{cm} in height) is surrounded by a 30-cm-thick radial graphite reflector and 10-cm-thick axial graphite reflectors positioned above and below the fuel.
The reactor is equipped with several irradiation facilities, including the Central Thimble, located at the core center, and the Lazy Susan facility, a specimen rack within the radial reflector. Both are well-suited for our NAA measurements. Detailed descriptions of the reactor's geometry and its irradiation facilities are available in references~\cite{AbsoluteFlux,BayesianSpectrum,FluxDistribution}.

The Lazy Susan facility is particularly advantageous due to its 40 irradiation channels, offering high flexibility in the number of samples that can be irradiated simultaneously. Each channel has a cylindrical shape with a diameter of approximately \qty{32}{mm}. We use  custom-designed polyethylene irradiation vessels  with an inner diameter of \qty{25}{mm} and a usable height of approximately \qty{100}{mm}, making them suitable for accommodating the PFA sample containers. 

Neutron fluxes at the Pavia TRIGA reactor are approximately \qty{1.7E13}{n/cm^2/s} in the Central Thimble and \qty{2.2E12}{n/cm^2/s} in the Lazy Susan facility~\cite{BayesianSpectrum}. Typical operations involve six hours of full-power irradiation per day, with our NAA campaigns scheduled accordingly.

%% file: 3.4.GeSpark.tex
As a result of the irradiation, neutron-rich nuclides are created, which return to stability with a \b-\ decay. These decays typically leave the daughter nucleus in an excited state, emitting characteristic \g-rays within a few picoseconds. Since these timescales are much shorter than the integration times of conventional measurement systems, the simultaneous emission of beta and gamma radiation is expected from neutron-activated nuclides. 
The \bg\ cascade signal serves as a discriminant for background events by utilizing two different detectors. The GeSparK system, developed at the  Radioactivity Laboratory of the University of Milano-Bicocca, is a low-background detector consisting of a liquid scintillation detector and an HPGe detector operating in temporal coincidence. 
Besides \bg\ coincidence measurements, GeSparK employs passive and active shielding to significantly reduce the background caused by environmental and cosmic radiation. Details of the system are provided in \cite{GS:2021,GSMuons}.
In particular, the scintillation detector includes a cylindrical PTFE container filled with Ultima Gold AB (Perkin Elmer) liquid scintillator as detection medium.
After post-irradiation treatments, the irradiated samples are mixed with this liquid scintillator for the \bg\ spectroscopy.
As described by the decay sequence in Section~\ref{S:NAA}, the \bg\ analysis focuses on identifying \pa(\np) decays to quantify \th(\u) concentrations in the irradiated samples. 

\pa\ has a half-life of about \qty{27}{days}, much longer than most interfering nuclides. Consequently, measurements to assess \th\ concentrations typically begin about one week after irradiation. The most intense \g-line in \pa\ decay occurs at \qty{312}{keV}, following a \b-\ decay with an endpoint energy of \qty{260}{keV}. By selecting \b-\ events in the GeSpark scintillation detector with energies up to \qty{260}{keV} in temporal coincidence with \g\ events at \qty{312}{keV} in the HPGe detector, spurious background is nearly eliminated. Under these conditions, the primary background source is the intrinsic background of the GeSparK system, dominated by cosmic muon showers~\cite{GSMuons}.

In contrast, \np\ has a half-life of about \qty{2.4}{days}, requiring measurements to assess \u\ concentrations while background from interfering nuclides is still present. Measurements typically span one week, allowing approximately 90\% of the $^{239}\textrm{Np}$ to decay. The most critical interfering nuclides are $^{24}\textrm{Na}$ and $^{82}\textrm{Br}$, which are common in the environment (Na) and organic compounds (Br), with half-lives comparable to that of $^{239}\textrm{Np}$. Sensitivity can be significantly improved by exploiting the unique nuclear decay characteristics of \np, which decays with a high branching ratio (40.5\%) to the 391~keV metastable level of $^{239}$Pu. This level has a half-life of about \qty{190}{ns}~\cite{Pu239Article}, which is much longer than the time resolution of the LS. We thus select signals featuring a couple of prompt-delayed events in the LS in coincidence with a gamma detected by the HPGe. This approach drastically reduces interference from other nuclides and intrinsic detector background. The technique used to analyze the prompt-delayed coincidences for the evaluation of \u\ contaminations is described in detail in \ref{S:np-239}.

%% file: 4.RecoveryEfficiency.tex
The evaluation of the process efficiency for both \u\ and \th, defined as the ratio of the mass of the target nuclides in the initial LS sample to their mass after processes 1 to 4 in Section~\ref{S:MeasStrategy}, is crucial for this analysis. 
To assess the efficiency, LAB samples were spiked with known amounts of $^{238}\textrm{U}$ and $^{232}\textrm{Th}$, and the final contaminant quantities were measured after the complete procedure. The contamination was introduced using certified standard solutions of U and Th in 1\% \ch{HNO3}, with a concentration of \qty{1000}{\micro g/g}, from Inorganic Ventures. Since these solutions are too concentrated for direct use and are not miscible with organic matrices, a stepwise dilution protocol was developed:
    \begin{itemize}
		\item Dilution 1:5: mix \qty{0.1}{mL} of U standard (\qty{1000}{\micro g/g}) and \qty{0.1}{mL} of Th standard (\qty{1000}{\micro g/g}) with \qty{0.3}{mL} of 1-2\% acidified water to obtain \qty{0.5}{mL} of a \qty{200}{\micro g/g} U/Th solution.
		\item Dilution 1:100: dilute \qty{0.1}{mL} of the above solution in \qty{9.9}{mL} of water, to prepare \qty{10}{mL} of a \qty{2}{\micro g/g} U/Th solution.
		\item Dilution 1:100: dilute \qty{0.1}{mL} of the above solution in \qty{9.9}{mL} of isopropyl alcohol to obtain a \qty{20}{ng/g} solution.
	\end{itemize}
The final dilution step uses isopropyl alcohol because, unlike aqueous solutions, it is miscible with LAB (a non-polar compound). From this solution, \qty{0.1}{mL} of the \qty{20}{ng/g} standard is added to \qty{100}{g} of LAB to achieve a contamination level of \qty{2E-11}{g/g} of \u\ and \th. This corresponds to a contamination of \qty{2}{ng} of \u\ and \qty{2}{ng} of \th\ in the LAB sample. 
The mixing of the standard solution with the LAB sample is performed on a magnetic stirrer, ensuring vigorous agitation for several minutes to achieve uniform dispersion of the nuclides within the LAB matrix.

Nine LAB samples, prepared using the described contamination procedure, were analyzed in four efficiency evaluation campaigns. To verify the initial contamination levels, \qty{0.1}{mL} of the same contaminated LAB solution was irradiated together with the samples, consistently showing uranium and thorium masses matching the expected value of \qty{2}{ng} in all analyzed cases. In parallel, the LAB samples were processed through steps 1 to 4 outlined in Section~\ref{S:MeasStrategy} and the final eluates were measured by \g\ spectroscopy to quantify the amounts of \u\ and \th.

\begin{figure}
		{\includegraphics[width=.48\textwidth]{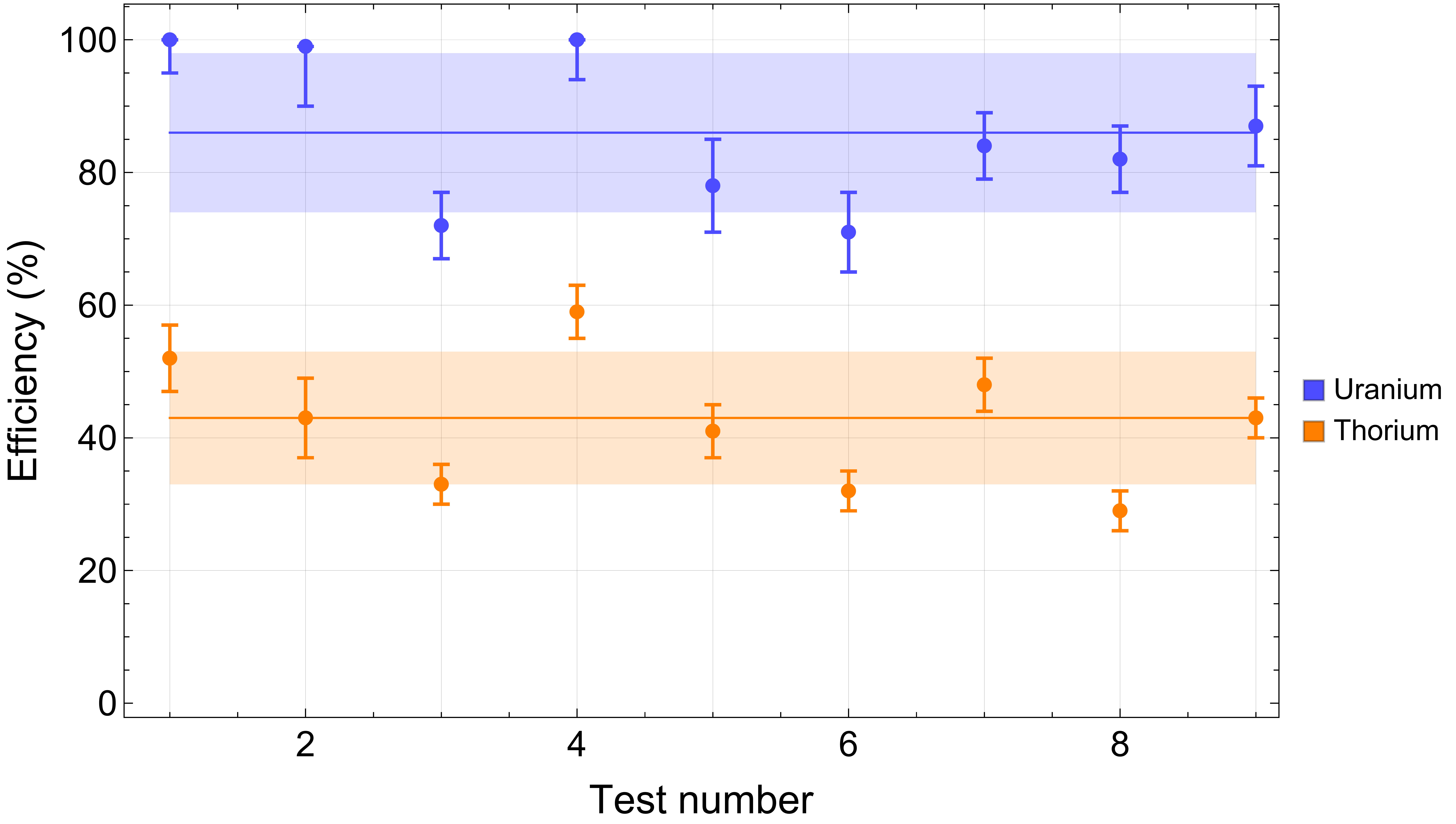}}
		\caption{Results of the total efficiency of the complete measurement procedure for uranium and thorium The coloured regions are the $1\sigma$ bands around the means.}
		\label{fig:TotalEfficiencies}	
	\end{figure}

The results are presented in Table~\ref{tab:CompleteEfficiencies} and illustrated in Figure~\ref{fig:TotalEfficiencies}.
    \begin{table}[t]
		\centering
		\begin{tabular}{ccc}
			\toprule
			Test number & U efficiency & Th efficiency \\
            & (\%) & (\%) \\
			\midrule
			1 & $100^{+0}_{-5}$& \num{52(5)} \\
			2 & $99^{+1}_{-9}$ & \num{43(6)} \\
			3 & \num{72(5)} & \num{33(3)} \\
			4 & $100^{+0}_{-6}$ & \num{59(4)} \\
            5 & \num{78(7)} & \num{41(4)} \\
            6 & \num{71(6)} & \num{32(3)} \\
            7 & \num{84(5)} & \num{48(4)} \\
            8 & \num{82(5)} & \num{29(3)} \\
            9 & \num{87(6)} & \num{46(3)} \\
			Mean  & \num{86(12)} & \num{43(10)} \\
			\bottomrule
		\end{tabular}
		\captionof{table}{Results of the total efficiency of the complete measurement procedure for uranium and thorium are summarized. The final row shows the average efficiency, with the error calculated as the standard deviation of the individual measurements.}
		\label{tab:CompleteEfficiencies}
    \end{table}
Uranium recovery demonstrates high efficiency, while thorium recovery is lower; however, both exhibit good reproducibility. The errors associated with the efficiencies are calculated as the standard deviation of individual measurements. The final efficiencies are  \qty{86(12)}{\%} for \u\ and to \qty{43(10)}{\%} for \th.

%% file: 5.BlankSamples.tex
\begin{table*}[ht]
    \centering
    \begin{tabular}{cccccc}
        \toprule
        Test & Mass (g) & \multicolumn{2}{c}{Mass (pg)} & \multicolumn{2}{c}{Concentration (\qty{E-15}{g/g})} \\
        &  & $^{238}\textrm{U}$ & $^{232}\textrm{Th}$ & $^{238}\textrm{U}$ & $^{232}\textrm{Th}$ \\
        \midrule
        1 & 228  & \num{1.8(0.3)} & \num{<22}  & \num{7.9(1.4)} & \num{<97}  \\
        2 & 228  & \num{2.2(0.2)} & \num{<12}* & \num{9.5(1.0)} & \num{<53}* \\
        3 & 232  & \num{4.9(0.9)} & \num{<23}  & \num{21(4)}    & \num{<100} \\
        4 & 46.6 & \num{7.2(1.2)} & \num{<25}  & \num{155(25)}  & \num{<540} \\
        \bottomrule
    \end{tabular}
    \captionof{table}{The results of the blank measurements applying the single washing protocol of the resin are presented. The second column shows the mass of the processed extracting solution, while the third and fourth columns list the resulting masses of $^{238}\textrm{U}$ and $^{232}\textrm{Th}$, estimated using the complete processing technique. The last two columns display the concentration of $^{238}\textrm{U}$ and $^{232}\textrm{Th}$, calculated as the ratio between the nuclide mass and the sample mass. *The sensitivity for $^{232}\textrm{Th}$ is approximately two times lower than in tests one and three, as this measurement was performed with a different HPGe detector with higher efficiency. This result is currently excluded from the evaluation of \th\ sensitivity.}
    \label{tab:Blank1}
\end{table*}

A blank sample is a \textit{dummy} sample with the same mass (or volume) as the actual sample to be measured. It undergoes the same procedures but is neither contaminated with the standards nor exposed to the substance being measured (in this case, the liquid scintillator matrix). The 5M \ch{HNO3} solution is processed using the extraction chromatography procedure with UTEVA resin. The eluate is then irradiated and further processed through extraction chromatography with TEVA resin, as previously described. Finally, the resulting eluate is measured with the GeSparK detector to determine the concentrations of $^{238}\textrm{U}$ and $^{232}\textrm{Th}$.
Blank test measurements primarily aim to assess the final sensitivity of the complete procedure while simultaneously evaluating potential \u\ and \th\ contaminations introduced during pre-irradiation treatments.

\begin{table}[ht]
    \centering
    \begin{tabular}{ccc}
        \toprule
        Test & Mass & $^{238}\textrm{U}$ mass \\
        & (g) & (pg) \\
        \midrule
        5 & 46.5 & \num{0.65(0.23)} \\
        6 & 48.3 & \num{0.18(0.12)} \\
        7 & - & \num{0.34(0.15)} \\
        \bottomrule
    \end{tabular}
    \captionof{table}{The results of the blank measurements using the new resin cleaning protocol based on double washing are presented. The second column shows the mass of the processed extracting solution and the final column reports the mass of $^{238}\textrm{U}$. The third measurement was performed without processing the extracting solution, simply eluting the column after cleaning.}
    \label{tab:Blank2}
\end{table}

Our initial blank tests involved only one preliminary washing step of the \uteva\ resin with \qty{50}{\mL} of washing solution. The results of these tests are summarized in Table~\ref{tab:Blank1}.
The first two measurements (tests one and two in the table) show consistent results for the residual contamination of \u\ and upper limits for \th. Considering that the extracting solution contains \ch{HNO3} at \qty{0.1}{M}, it is possible to estimate the concentration of $^{238}\textrm{U}$ in the concentrated nitric acid (\qty{15}{M} \ch{HNO3}), assuming that the entire contamination of the blank originates from it. However, the calculated concentrations are inconsistent with both the ICP-MS measurements of the nitric acid performed initially (see Section~\ref{S:Labware}) and the certified value provided by the manufacturer.
Based on these inconsistencies, we conducted two additional blank measurements: one using the same solution mass and another with a significantly lower mass, to determine whether the contamination originates from the acid or the resin. If the acid is the source of contamination, the resulting concentrations should scale as the sample mass. Conversely, if the contamination arises from the resin, we would expect a nearly constant amount of uranium (U mass) released, independent of the solution mass.
The results of tests three and four in Table~\ref{tab:Blank1} confirmed the latter hypothesis: the mass of U in the final sample is independent of the solution mass, indicating a constant release from the resin.

Under these conditions, the expected sensitivity for the measurement of \qty{1}{L} of liquid scintillator (\qty{860}{g}, considering the density of 0.86 g/cm$^3$) is \qty{9.7E-15}{g/g} for $^{238}\textrm{U}$, calculated as described in \ref{S:lod} (Eq.~\ref{eq:LD}) from the standard deviations of the \u\ mass released in the different blank measurements of Table~\ref{tab:Blank1}. 
For $^{232}\textrm{Th}$, the sensitivity is limited by the intrinsic background of the GeSparK detector. Using the GeSparK configuration adopted in our initial tests (Table~\ref{tab:Blank1}) with the preliminary implementation of the improved veto system described in~\cite{GSMuons}, a sensitivity of \qty{26E-15}{g/g} was achieved for the measurement of \qty{1}{L} of liquid scintillator. After the full upgrade of the muon veto, a subsequent one-month measurement reached a sensitivity of \qty{21E-15}{g/g} for the same LS volume (Table~\ref{tab:FinalSensitivity}), calculated with Eq.~\ref{eq:LD4}.

To enhance uranium sensitivity, we tested a modified cleaning protocol involving two consecutive cleaning steps with an intermediate conditioning phase, as described in Section~\ref{S:MeasStrategy}, to reduce uranium release from the resin. The results of the new blank measurements are presented in the first two rows of Table~\ref{tab:Blank2}. The amount of $^{238}\textrm{U}$ released during the elution phase was reduced by approximately one order of magnitude. A third test, performed without processing any extracting solution -- eluting the column resin immediately after the cleaning -- yielded results consistent with the first two tests. This consistency confirms the hypothesis that the resin itself is the source of the contamination.
\begin{figure*}[h]
    \centering
    \subfloat[\label{fig:DTPlot1}Blank test 4]
    {\includegraphics[width=.48\textwidth]{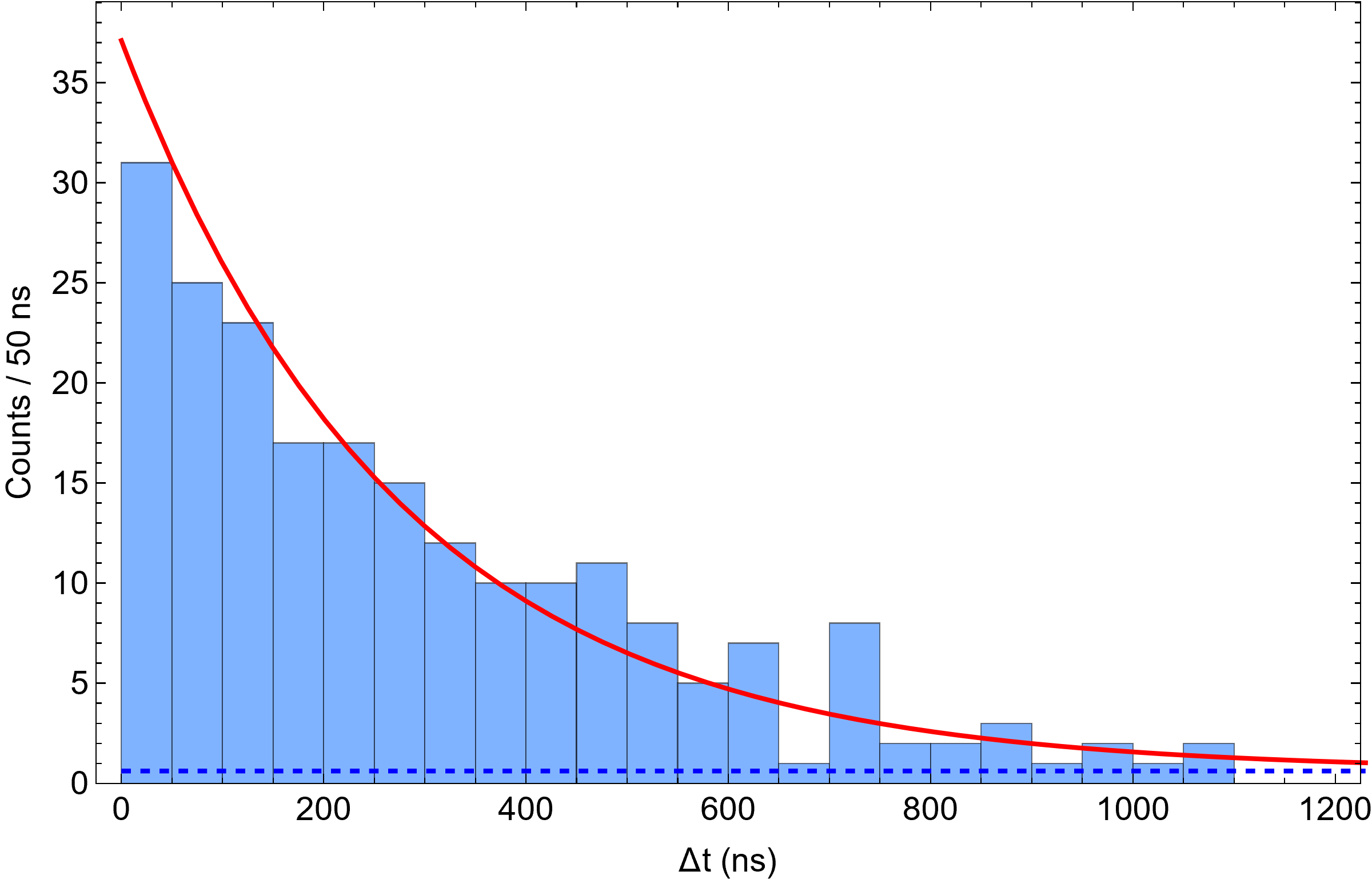}}
    \subfloat[\label{fig:DTPlot2}Blank test 5]
    {\includegraphics[width=.48\textwidth]{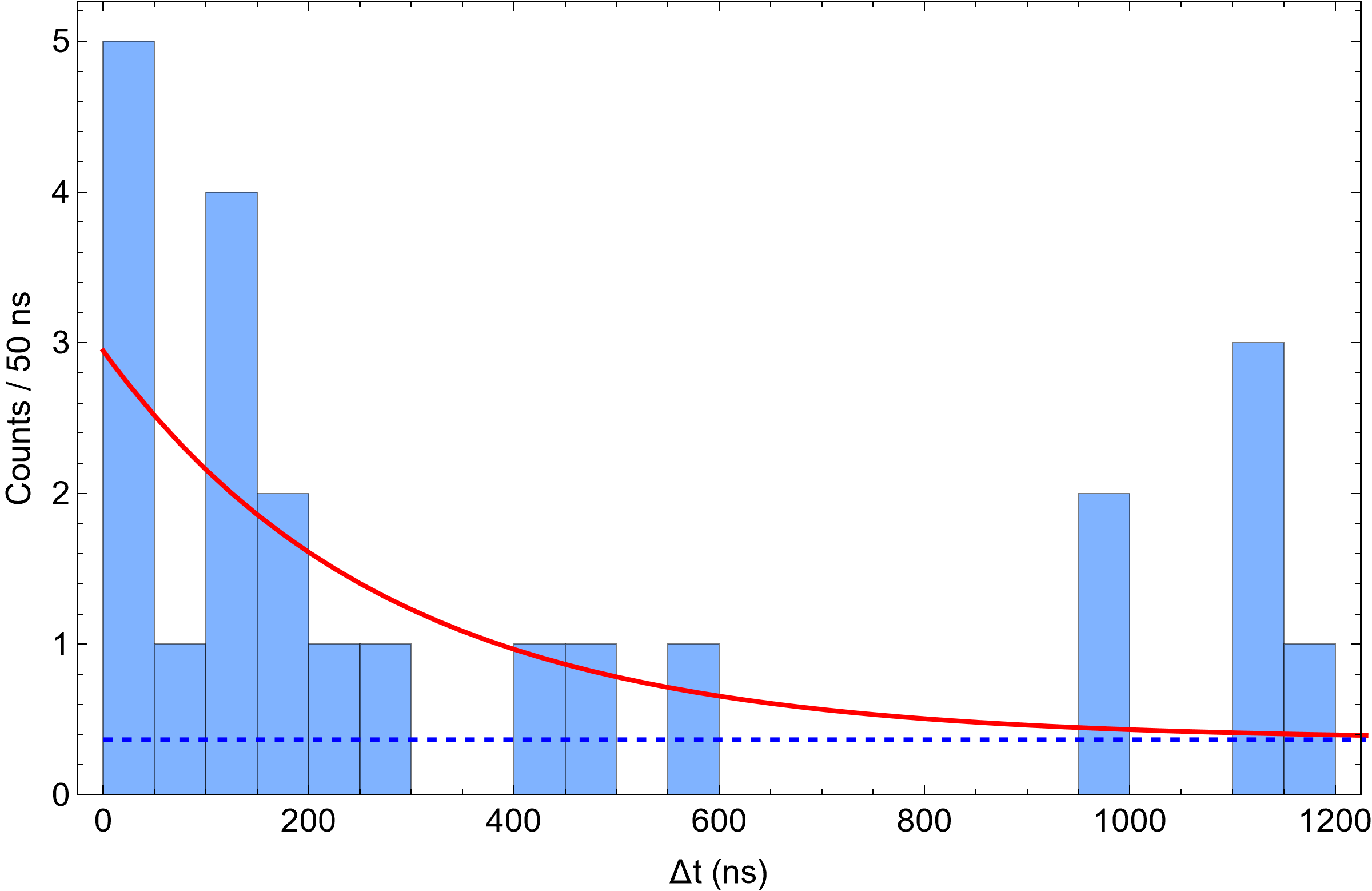}}
    \caption{Time distribution of $^{239}\textrm{Pu}$ delayed events in two blank samples with one washing step (\ref{fig:DTPlot1}) and two washing steps (\ref{fig:DTPlot2}). The plots also show the PDFs obtained from the JAGS fit (see \ref{S:np-239}), where the dashed blue line represents the background component and the solid red line represents the total distribution. The observed exponential decay is consistent with the half-life of the \np\ metastable state. The volume of the extracting solution, the irradiation time, and the measurement duration are identical for both samples.}
    \label{fig:UAnalysis}
\end{figure*}
Fig.\,\ref{fig:UAnalysis} presents the time distributions of events recorded in the GeSparK scintillation detector for test 4 in Table\,\ref{tab:Blank1} and test 5 in Table\,\ref{tab:Blank2}. The effectiveness of the two-step washing process is evident from the significantly lower count rate observed in the second test.

\begin{figure*}[h]
    \centering
    \subfloat[\label{fig:ICPMSU}] 
    {\includegraphics[width=.48\textwidth]{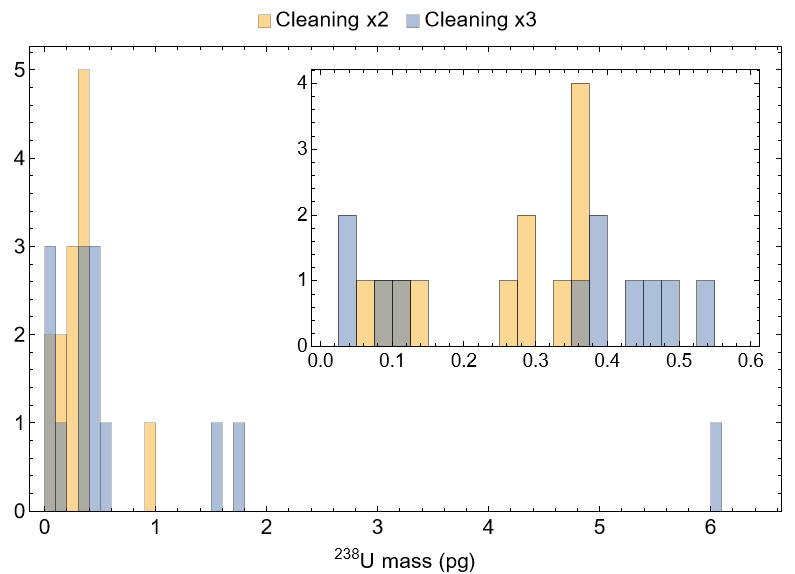}}
    \subfloat[\label{fig:ICPMSTh}]
    {\includegraphics[width=.48\textwidth]{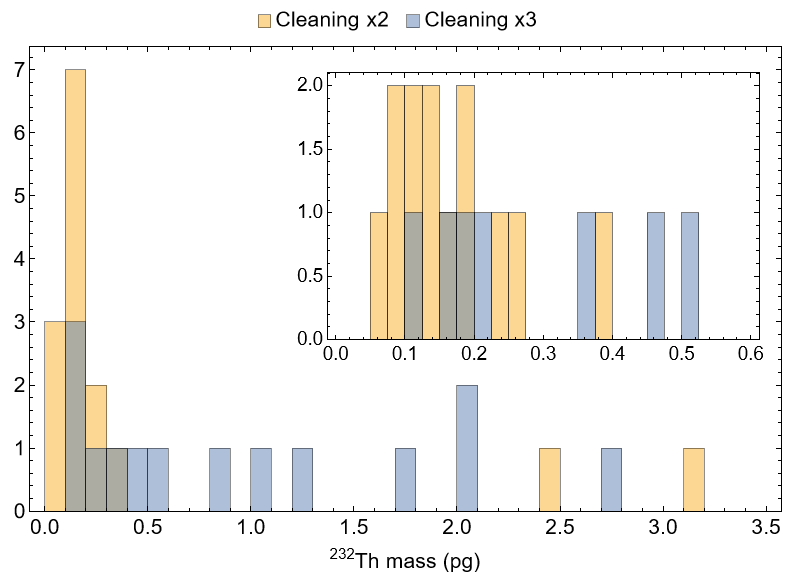}}
    \caption{Masses of U and Th released by UTEVA resins, measured by ICP-MS. The elution was performed with \qty{15}{mL} of \qty{0.02}{M} \ch{HNO3} following the double and triple cleaning protocols. In the upper right boxes, the distributions of values below \qty{0.6}{pg} are displayed, which were used to compute the mean release values: $m_\textrm{U}=\qty{0.29(0.17)}{pg}$ and $m_\textrm{Th}=\qty{0.21(0.13)}{pg}$.}
    \label{fig:ICPMSResins}
\end{figure*}

Using ICP-MS measurements at LNGS, we then tested several \uteva\ resins to gain a thorough understanding of the background, applying the double-washing cleaning protocol and also evaluating a triple-washing protocol to explore potential improvements. The results are summarized in the two plots in Figure~\ref{fig:ICPMSResins}, which show the uranium and thorium release after the double and triple washing procedures.
As shown in the plots, the results are generally consistent, with only a few outliers. Removing these outliers, as indicated in the zoomed-in boxes, is justified by the fact that, based on the ICP-MS pre-screening, we know the actual contamination released by these resins and we will not use the most contaminated ones for our samples. Another key outcome of the ICP-MS measurements is that adding a third washing step does not improve the results; in fact, there are more outliers than in the two-washing step case.

By combining all measurements from the double and triple cleaning protocols obtained through NAA and ICP-MS below a chosen threshold of \qty{0.6}{pg} -- a cut that retains approximately 75\% of the measured resins -- we calculated the mean value and standard deviation for both $^{238}\textrm{U}$ and $^{232}\textrm{Th}$ release from the \uteva\ resins. Using these results and Eq.~\ref{eq:LD}, we assessed the final sensitivity of the method for $^{238}\textrm{U}$ and the maximum achievable sensitivity for \th\ (in case of no limiting background from the GeSpark detector). The results are summarized in Table~\ref{tab:ResinRelease}.

\begin{table}[t]
    \centering
    \begin{tabular}{ccc}
        \toprule
        & Mass (pg) & Sensitivity \\
        &&(\qty{E-15}{g/g}) \\
        \midrule
        $^{238}\textrm{U}$ & \num{0.29(0.17)} & \num{0.65} \\
        $^{232}\textrm{Th}$ & \num{0.21(0.13)} & \num{0.50} \\
        \bottomrule
    \end{tabular}
    \captionof{table}{Sensitivity limitation due to \uteva\ resin background release. The second column shows the mean release of $^{238}\textrm{U}$ and $^{232}\textrm{Th}$ from the resins, along with their standard deviation, calculated from all NAA and ICP-MS measurements below the \qty{0.6}{pg} threshold. The third column indicates the computed sensitivity for measuring \qty{1}{L} (\qty{860}{g}) of liquid scintillator, assuming it is solely limited by resin contamination.}
    \label{tab:ResinRelease}
\end{table}

\begin{figure*}
    \centering
    \includegraphics[width=.95\textwidth]{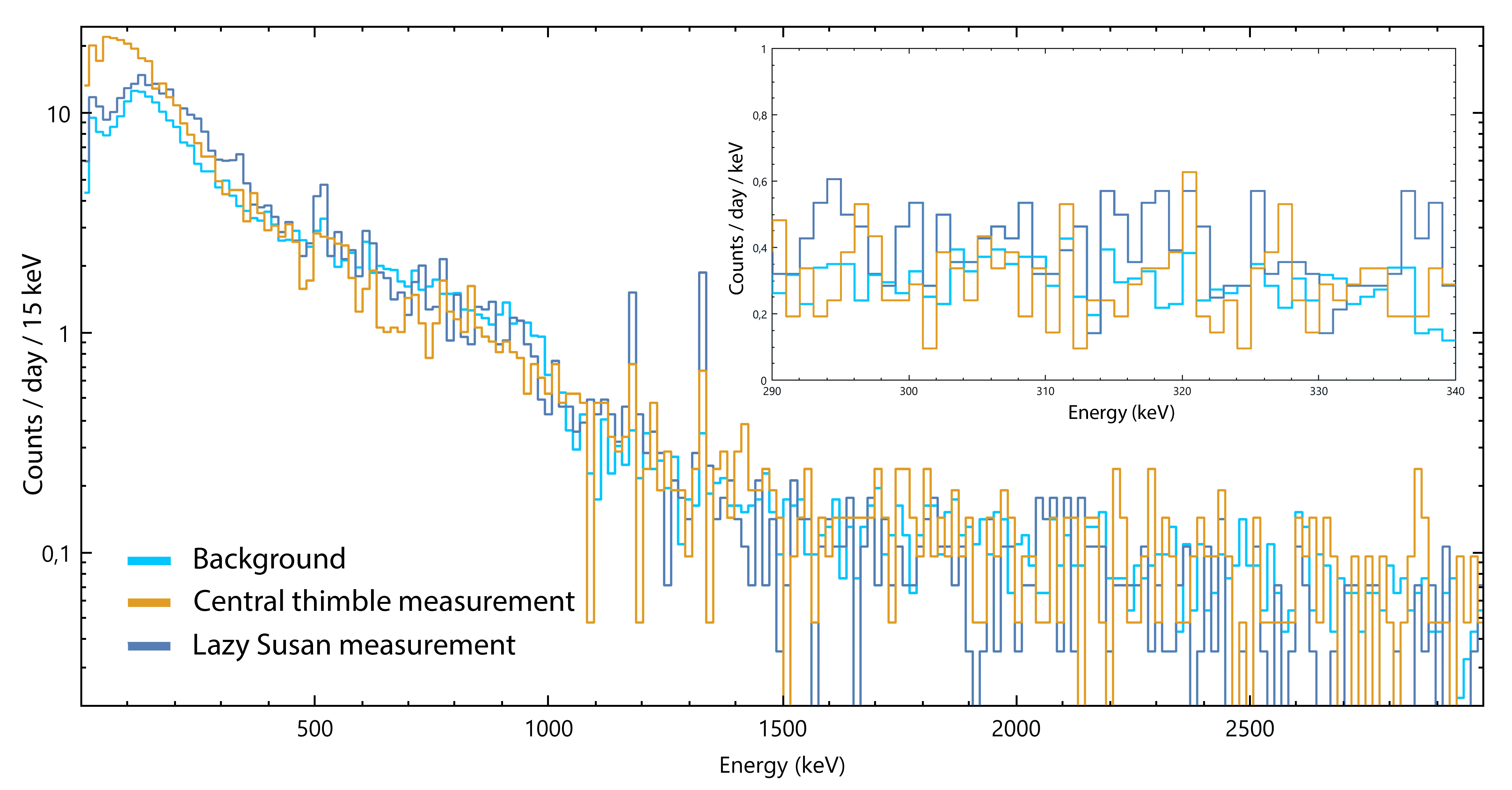}
    \caption{HPGe detector spectra in coincidence with a signal above \qty{60}{keV} in the LS detector for the detector background (light blue), Lazy Susan measurement (blue), and Central Thimble measurement (orange). The upper-right inset highlights the region of interest around the \qty{312}{\keV} \pa\ gamma line. The measurement durations are: 91 days for the background, 28 days for the Lazy Susan measurement, and 21 days for the Central Thimble measurement.}
    \label{fig:HPGeSpectra}
\end{figure*}

The long half-life of \pa\ and the impossibility of applying delayed coincidence techniques to suppress background strongly limit the sensitivity to \th\, which remains significantly worse than the achievable performance, reported in Table \ref{tab:ResinRelease}. To improve it, we evaluated the possibility of irradiating the sample with a much higher neutron fluence by exploiting the Central Thimble of the TRIGA reactor. As described in Section~\ref{S:Triga}, this channel provides a higher thermal neutron flux than the Lazy Susan channel, but also a higher fast neutron flux. Preliminary tests were performed to verify the resistance of the PFA vials to the high neutron flux, particularly for the fast component of the spectrum, yielding satisfactory results.

Based on these tests, we conducted a dedicated blank measurement using the TRIGA Central Thimble irradiation channel to evaluate the achievable \th\ sensitivity, obtaining a minimum detectable concentration of \qty{1.9E-15}{g/g} for \qty{1}{L} of LS, calculated with Eq.~\ref{eq:LD4}. The final sensitivities of our analysis procedure for \u\ and \th\ are summarized in Table \ref{tab:FinalSensitivity}. 
Fig.\,\ref{fig:HPGeSpectra} presents the HPGe spectrum acquired with GeSparK for the Central Thimble measurement, compared to those from the Lazy Susan irradiation and the intrinsic GeSparK background. The impact of the GeSparK background limitation is clearly visible. To calculate the \th\ sensitivity reported in Table\,\ref{tab:FinalSensitivity}, the background counts were integrated over two \qty{20}{keV}-wide energy intervals on either side of the expected \qty{312}{keV} peak from \pa, ensuring that the well-established background evaluation described in \ref{S:lod} remains valid.

\begin{table}[t]
    \centering
    \begin{tabular}{ccc}
        \toprule
         & TRIGA channel & Sensitivity \\
         && (\qty{E-15}{g/g}) \\
        \midrule
        $^{238}\textrm{U}$ & Lazy Susan & \num{0.65} \\
        $^{232}\textrm{Th}$ & Lazy Susan & \num{21} \\
        $^{232}\textrm{Th}$ & Central Thimble & \num{1.9} \\
        \bottomrule
    \end{tabular}
    \captionof{table}{Our current sensitivity for measuring trace concentrations of \u\ and \th\ in \qty{1}{L} (\qty{860}{g}) of liquid scintillator. The measurement times are one week for \u\ and \qty{30}{days} for \th.}
    \label{tab:FinalSensitivity}
\end{table}

%% file: Conclusions.tex
In this paper, we presented our measurement strategy to determine trace concentrations of \u\ and \th\ at sensitivities reaching the \ppq\ level. The method combines neutron activation analysis with radiochemical techniques, involving pre-concentration and purification of the sample using acidified water and two stages of solid-phase extraction -- one prior to and one following neutron irradiation. Each sample is subsequently measured using a custom \bg\ detector, GeSpark, which enables background discrimination through coincidence spectroscopy. Benefiting from the favorable nuclear decay scheme of \np, our current \u\ sensitivity is \qty{0.65E-15}{g/g} limited only by uranium released from the \uteva\ resin during the pre-irradiation step. For \th, the sensitivity is presently limited by the intrinsic background of the GeSpark detector, achieving \qty{1.9E-15}{g/g} after neutron irradiation in the Central Thimble of the TRIGA nuclear reactor in Pavia. These sensitivities rank among the best currently achieved worldwide.

Potential improvements to our approach include replacing the pre-irradiation extraction chromatography step with the use of a vacuum oven to concentrate the sample. This substitution would facilitate handling larger masses of liquid scintillator, which would only impact the liquid extraction step, and streamline the whole process. Increasing the initial LS mass would proportionally enhance the measurement sensitivity for both \u\ and \th. For \th, which is not yet limited by resin contamination, this improvement would be particularly significant.
Moreover, the use of a different type of resin -- e.g., \truR, which is suitable for actinide separation -- could improve the efficiency of \th\ measurements, currently about half that of \u. The typically higher background levels of \tru\ compared to \uteva\ would not pose an issue since it would be used exclusively after neutron irradiation, when there is no risk of sample contamination.
Another step forward in \th\ sensitivity could involve deploying a new \bg\ detection system utilizing an HPGe detector with improved detection efficiency, as demonstrated with a temporary prototype (test 2 in Table~\ref{tab:Blank1}).

The method outlined in this paper has been applied during the commissioning phase of the JUNO detector to measure LS samples following the various LS purification steps~\cite{juno-ppnp,impianti}, ensuring compliance with the radiopurity requirements of JUNO.

%% file: appendix_LOD.tex
\section{Experimental sensitivity calculation}
\label{S:lod}

In this appendix, we summarize the statistical methods applied to analyze the $\gamma$-spectroscopy measurements of samples with ultra-low contamination, where signals are difficult to be distinguished from background (or blank) fluctuations.
According to the seminal paper published by Currie in 1968~\cite{Currie:1968}, the \textit{critical limit} $L_C$ is defined as the threshold above which a measurement result indicates a detected non-zero signal with a certain confidence level (typically 95\%).

This limit is distinct from the \textit{detection limit} $L_D$, which is defined as the true value of the signal for which the probability of obtaining a result lower than $L_C$ is equal to a specified value (typically 5\%), considered acceptable. The detection limit reflects the capability of an analytical method to detect the presence of a signal and is, therefore, sometimes referred to as \textit{sensitivity}.

In this analysis, we faced two different situations in the search for \u\ and \th\ contaminations in the blank samples.
Indeed, when analyzing blank samples, we found non-zero signals for \u\ (i.e. results above $L_C$), whereas no peaks from \th\ contaminations were observed in the spectra.

To determine the experimental sensitivity of our measurement technique, we calculated the \textit{detection limits} $L_D$ using Currie's formula, assuming a \textit{well-known} background (or blank signal) with a probability density function (PDF) that can be approximated by a Gaussian distribution.
\begin{equation}
\label{eq:LD}
    L_D=2\,L_C=3.29 \, \sigma_B
\end{equation}
where $\sigma_B$ is the standard deviation of background (or blank) PDF.
In particular, to assess the detection limit for \u\ contaminations, we calculated $\sigma_B$ as the standard deviation of the non-zero \u\ masses measured in the blank samples.

The \th\ detection limit was instead evaluated by setting $\sigma_B$ equal to the \th\ mass that would produce a number of signal counts in the spectrum equal to $\sqrt{\mu_B}$, where $\mu_B$ represents the average background counts expected within a 1.25 FWHM-wide range centered on the peak energy. In this context, we emphasize that $\mu_B$ can be determined with good precision (thus supporting the 'well-known' background hypothesis), as we characterize the background counting rate per unit energy by analyzing a relatively wide energy range on both sides of the peak region~\cite{DEGEER2004151}.

In conclusion, when a sample is analyzed to detect \u\ and \th\ contaminations, the results are determined as follows.
\begin{itemize}
    \item If a non-zero signal is observed in the blank measurements, it can be subtracted from the sample measurement to compute the net contamination of the sample:
\begin{equation}
\label{eq:LD2}
	C = \frac{m_\textrm{meas}-m_\textrm{blank}}{M_\textrm{sample}}
\end{equation}
where $m_\textrm{meas}$ is the \u/\th\ mass determined from the sample measurement, $m_\textrm{blank}$ is the mean \u/\th\ in the blank measurements, and $M_\textrm{sample}$ is the sample mass. 
If $m_\textrm{meas}-m_\textrm{blank} > L_C$ ($L_C=1.645\,\sigma_B$), we conclude that the sample contains additional contamination compared to the blank. Otherwise, we report an upper limit (95\% C.L.) on the concentration:
\begin{equation}
\label{eq:LD3}
    C<\frac{3.29 \, \sigma_B}{M_\textrm{sample}}
\end{equation}
We emphasize that applying this procedure requires conducting a sufficient number of blank measurements to verify repeatability and accurately determine $\sigma_B$ of their PDF.

\item When the blank signal is not visible because hidden by the detector background (as was the case of \th\ measurements in this work), the concentration sensitivities are determined based on the number of background events ($\mu_B$), without subtracting any blank contribution. In particular, when $\mu_B$ in the signal range is too low for the Gaussian approximation to be valid, we use the more general Currie’s formula to calculate the detection limit of signal counts:
\begin{equation}
\label{eq:LD4}
    L_D = 2.71+3.29\sqrt{\mu_B}
\end{equation}
which we use to determine the upper limit on the contaminant concentration.
\end{itemize}

%% file: appendix_Np-239.tex
\section{Delayed coincidence techniques for $\textrm{U}$ determination}
\label{S:np-239}

The standard NAA technique for determining $^{238}\textrm{U}$ relies on measuring the \qty{106}{\keV} gamma ray emitted in the decay of $^{239}\textrm{Np}$, as this provides the highest branching ratio and sensitivity. By comparing the measured counts for the sample and the standard, the concentration of $^{238}\textrm{U}$ in the sample is determined using the NAA equation.

A key limitation to the sensitivity of $^{239}\textrm{Np}$ measurements arises from background events around \qty{106}{\keV} caused by other activated nuclides, such as $^{24}\textrm{Na}$ and $^{82}\textrm{Br}$, and by the detector’s intrinsic background. To address this, we implemented an advanced ``delayed coincidence technique'', inspired by \cite{Goldbrunner1997}, that significantly reduces background interference. This method leverages a metastable level in the $^{239}\textrm{Pu}$ decay scheme, acting as a unique marker for $^{239}\textrm{Np}$ decay.

\begin{figure}[t]
    \begin{center}
        \includegraphics[width=0.45\textwidth]{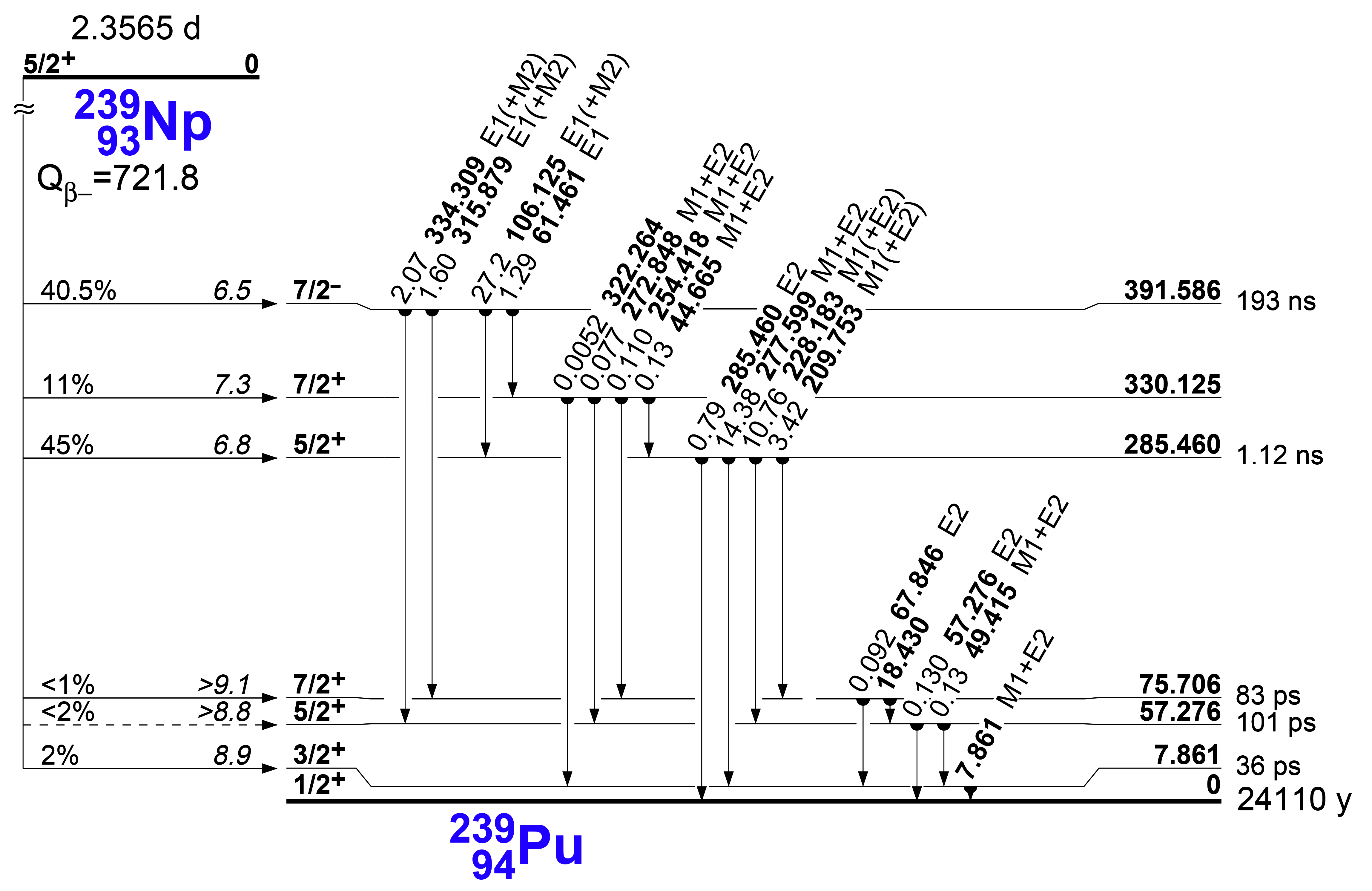}
        \caption{Simplified nuclear level scheme of $^{239}\textrm{Pu}$\cite{TOI}. The main transitions associated with the de-excitation of the metastable level are shown. All the most intense gamma rays emitted in the decay of this metastable state lie below \SI{300}{keV}, which is relevant for the identification of delayed-coincidence signatures.}
        \label{fig:Np239scheme}
    \end{center}
\end{figure}

\begin{figure}[t]
    \begin{center}
        \includegraphics[width=0.45\textwidth]{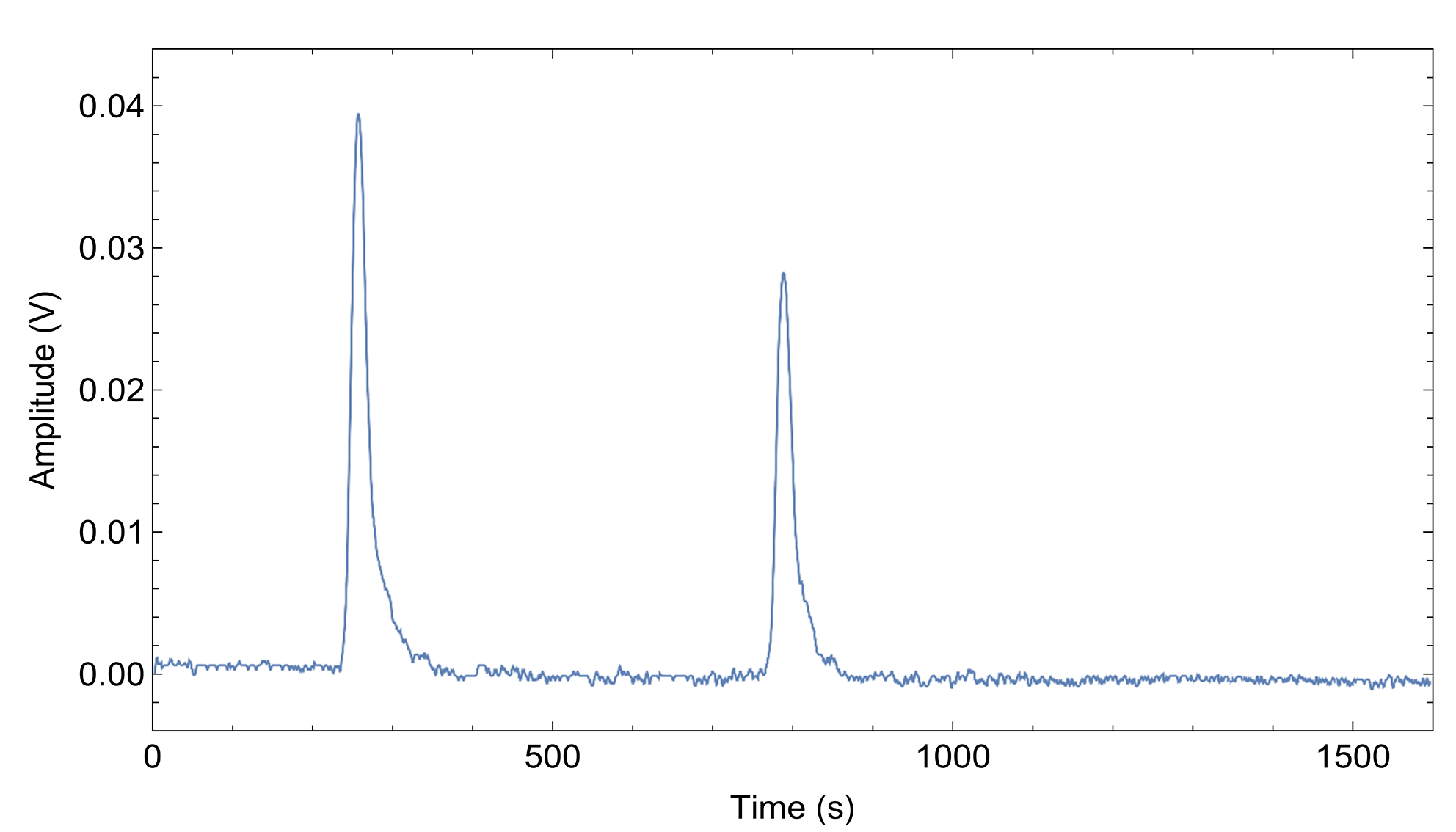}
        \caption{Example of delayed coincidence signals of $^{239}\textrm{Np}$ decay as acquired by the GeSparK detector.
        Example of delayed-coincidence signals from the decay of $^{239}\textrm{Np}$, as recorded by the GeSparK detector. The first pulse corresponds to the $\beta^-$ transition populating the metastable level of $^{239}\textrm{Pu}$, while the second pulse is produced by the internal conversion (or gamma/X-ray) associated with the de-excitation of the metastable state. This latter pulse is in time coincidence with a gamma/X-ray detected by the HPGe detector.}
        \label{fig:GSSignals}
    \end{center}
\end{figure}

The $^{239}\textrm{Np}$ beta decay (40.5\% branching ratio) populates a metastable level at \qty{391.6}{\keV}, followed by internal conversion (IC) or gamma transitions with a characteristic time delay (Figure~\ref{fig:Np239scheme}). The GeSparK detector, thanks to the excellent time resolution of the liquid scintillator, can simultaneously detect the beta decay and the subsequent delayed IC signal, allowing for a precise measurement of the time interval between the two events (Figure~\ref{fig:GSSignals}). The LS signals are acquired in coincidence with the associated gamma/X-ray detected by the HPGe detector.
This coincidence pattern provides robust markers for identifying $^{239}\textrm{Np}$ decays while rejecting most accidental coincidences caused by interfering nuclides or cosmic muons.

To identify valid events, a custom algorithm processes the GeSparK detector data to extract time differences, pulse amplitudes, and gamma energies. Several selection cuts are applied: the time delay must exceed the LS pulse width to avoid misreconstruction of the deposited energy; the LS pulse energies must match the expected values; and the HPGe gamma energy must fall below \qty{300}{\keV} (see Figure~\ref{fig:Np239scheme}). 

Signal events follow an exponential decay distribution with a characteristic half-life of \qty{190.2}{ns} \cite{Pu239Article}, while background events are modeled as uniform distributions within the time window.
Since the detected events include both signal and background, the overall distribution is the weighted sum of their respective distributions, with the weight determined by the expected fraction of signal events relative to the total ($c$ parameter). The analytical expression is given by:

\begin{equation}
    \mathcal{D} = c \cdot \frac{ \frac{1}{\tau}e^{-\frac{t}{\tau}} }{\int_{0}^{\Delta t} \frac{1}{\tau}e^{-\frac{t}{\tau}}dt} + (1-c)\cdot \frac{1}{\Delta t}
    \label{eq:Pu239GenericDistibution}
\end{equation}

This equation accounts for the truncation of the exponential distribution at time $\Delta t$ due to the finite acquisition window. The truncation is incorporated in two ways: (1) the denominator integral normalizes the exponential distribution over $[0, \Delta t]$, and (2) the uniform distribution is normalized over the same interval. Here, the parameter $c$ represents the exact fraction of signal events among all observed events. The GeSparK detector records the LS signal for \qty{1600}{\ns} per trigger, with a \qty{250}{\ns} pre-trigger, leaving a \qty{1350}{\ns} window for the second pulse. A \qty{120}{\ns} threshold excludes overlapping pulses, resulting in an effective signal window $\Delta t$ of \qty{1230}{\ns}, which is the value used in equation~\ref{eq:Pu239GenericDistibution}.

We perform a Bayesian unbinned fit of the time-difference distribution of delayed-coincident events using the model of Eq~\ref{eq:Pu239GenericDistibution}. The free parameter in the fit is the signal fraction c, which is assigned a uniform prior over the interval [0,1]. By analyzing the posterior distribution of c, we either extract its mean and standard deviation or, if the signal fraction is compatible with zero, determine its 90\% credible upper limit.

we employ a Bayesian analysis based on Markov Chain Monte Carlo (MCMC) using JAGS \cite{JAGS}, integrated into the GeSparK software. This approach accounts for statistical fluctuations in the total event count and provides posterior distributions for signal events, enabling precise determination of $^{239}\textrm{Np}$ activity in the sample and standard.

The same delayed coincidence analysis is applied to both sample and standard to perform the usual comparative NAA analysis. For the standard, where event counts are much higher, a binned chi-square minimization of the distribution $ a\cdot\frac{1}{\tau} e^{-\frac{t}{\tau}} + c $ is used instead of the unbinned MCMC approach. This ensures accurate, high-sensitivity measurements of $^{238}\textrm{U}$ using neutron activation analysis.